\documentclass[a4paper,11pt]{article}
\pdfoutput=1 

\usepackage{jcappub} 

\usepackage{multirow}

\title{A Micromegas-based low-background x-ray detector coupled to a slumped-glass telescope for axion research}

\author[a,1]{F.~Aznar,\note{Present address: Centro Universitario de la Defensa, Universidad de Zaragoza. Ctra. de Huesca s/n, 50090, Zaragoza, Spain}}
\author[a]{J.~Castel,}
\author[b]{F.~E.~Christensen,}
\author[a]{T.~Dafni,}
\author[c]{T.~A.~Decker,}
\author[d]{E.~Ferrer-Ribas,}
\author[a]{J.~A.~Garcia,}
\author[d]{I.~Giomataris,}
\author[a]{J.~G.~Gracia,}
\author[e]{C.~J. Hailey,}
\author[c]{R.~M.~Hill,}
\author[a]{F.~J.~Iguaz,}
\author[a]{I.~G.~Irastorza,}
\author[b]{A.~C~Jakobsen,}
\author[a]{G.~Luzon,}
\author[a]{H.~Mirallas,}
\author[d]{T.~Papaevangelou,}
\author[c]{M.~J.~Pivovaroff,}
\author[c]{J.~Ruz,}
\author[f]{T.~Vafeiadis}
\author[c]{and J.~K.~Vogel}

\affiliation[a]{Grupo de F\'{\i}sica Nuclear y Astropart\'{\i}culas, Universidad de Zaragoza\\ C/ P. Cerbuna 12 50009, Zaragoza, Spain}
\affiliation[b]{DTU Space, National Space Institute, Technical University of Denmark\\Elektrovej 327, DK-2800 Lyngby, Denmark}
\affiliation[c]{Lawrence Livermore National Laboratory\\ Livermore, CA 94550, USA}
\affiliation[d]{CEA, IRFU, Centre d'\'Etudes Nucl\'eaires de Saclay \\ Gif-sur-Yvette, France}
\affiliation[e]{Physics Department and Columbia Astrophysics Laboratory, Columbia University\\ New York, NY 10027, USA}
\affiliation[f]{Aristotle University of Thessaloniki\\ 54124 Thessaloniki, Greece}

\emailAdd{faznar@unizar.es}
\emailAdd{jfcastel@unizar.es}
\emailAdd{finn@space.dtu.dk}
\emailAdd{tdafni@unizar.es}
\emailAdd{decker4@llnl.gov}
\emailAdd{esther.ferrer-ribas@cea.fr}
\emailAdd{jagarpas@unizar.es}
\emailAdd{ioanis.giomataris@cern.ch}
\emailAdd{jgraciag@unizar.es}
\emailAdd{chuckh@astro.columbia.edu}
\emailAdd{runnemrand@yahoo.com}
\emailAdd{iguaz@unizar.es}
\emailAdd{igor.irastorza@cern.ch}
\emailAdd{jakobsen@space.dtu.dk}
\emailAdd{luzon@unizar.es}
\emailAdd{mirallas@unizar.es}
\emailAdd{thomas.papaevangelou@cea.fr}
\emailAdd{pivovaroff1@llnl.gov}
\emailAdd{ruzarmendari1@llnl.gov}
\emailAdd{Theodoros.Vafeiadis@cern.ch}
\emailAdd{vogel9@llnl.gov}
 
\abstract{We report on the design, construction and operation of a low background x-ray detection line composed of a shielded Micromegas (micromesh gaseous structure) detector of the microbulk technique. The detector is made from radiopure materials and is placed at the focal point of a $\sim$~5 cm diameter, 1.3 m focal-length, cone-approximation Wolter I x-ray telescope (XRT) comprised of thermally-formed (or ``slumped'') glass substrates deposited with multilayer coatings. The system has been conceived as a technological pathfinder for the future International Axion Observatory (IAXO), as it combines two of the techniques (optic and detector) proposed in the conceptual design of the project. It is innovative for two reasons:  it is the first time an x-ray optic has been designed and fabricated specifically for axion research, and the first time a Micromegas detector has been operated with an x-ray optic. The line has been installed at one end of the CERN Axion Solar Telescope (CAST) magnet and is currently looking for solar axions.  The combination of the XRT and Micromegas detector provides the best signal-to-noise ratio obtained so far by any detection system of the CAST experiment with a background rate of 5.4$\times$10$^{-3}\;$counts per hour in the energy region-of-interest and signal spot area.}

\begin{document}
\maketitle
\flushbottom

\section{Introduction}
Axions appear in very well motivated extensions of the Standard Model (SM) including the Peccei-Quinn mechanism proposed to solve the long-standing strong-CP problem~\cite{Peccei:1977hh,Peccei:1977ur,Weinberg:1977ma,Wilczek:1977pj}. They are pseudoscalar, very light particles that generically couple with two photons~\cite{Turner:1989vc}. Therefore, axions can convert into photons (and vice versa) in the presence of electromagnetic fields, a process that is sometimes referred to as Primakoff effect \cite{Primakoff:1951}, in analogy with the pion-photon conversion. They can be produced non-relativistically in the early Universe~\cite{Sikivie:2006ni,Wantz:2009it}, and thus they are also a favoured candidate to solve the Dark Matter (DM) problem. Like the case of the Weakly Interacting Massive Particles (WIMPs) in supersymmetric theories, axions are appealing due to the fact that they are not an \textit{ad hoc} solution to the DM problem. While WIMPs and axions could each account on their own for all of the DM, several theories~\cite{Baer:2011uz,Bae:2013pxa} favour the possibility of a mixed WIMP-axion DM. In addition, several extensions of the SM (e.g., string theory~\cite{Arvanitaki:2009fg,Cicoli:2012sz,Ringwald:2012cu}) predict more generic axion-like particles (ALPs). These could also be produced in the early Universe and be part of the DM~\cite{Arias:2012az}. Both axions and ALPs are repeatedly invoked to explain a number of poorly understood astrophysical observations (see e.g.~\cite{Irastorza:1567109} and references therein). 

Diverse complementary experimental approaches~\cite{ANDP:ANDP201300727} are being employed in the search for axions and ALPs. In particular, axion helioscopes~\cite{Sikivie:1983ip} look for axions emitted by the Sun, and therefore do not rely on the assumption of axions being the DM. The emission of axions by the solar core is a robust prediction involving well known solar physics and the Primakoff conversion of plasma photons into axions. Solar axions have $\sim$keV energies and in strong laboratory magnetic fields can convert back into detectable x-ray photons. The basic layout of an axion helioscope thus requires a powerful magnet coupled to one or more x-ray detectors. When the magnet is aligned with the Sun, an excess of x-rays at the exit of the magnet is expected, over the background measured at non-alignment periods.

Currently, the most powerful axion helioscope is the CERN Axion Solar Telescope (CAST), now in operation at CERN for more than a decade. The experiment uses a Large Hadron Collider (LHC) dipole prototype magnet with a magnetic field of up to 9~T over a length of 9.3~m and an aperture of $2\times15~$cm$^2$~\cite{Zioutas:1998cc}, that is able to follow the Sun for $\sim3$ hours per day using an elevation and azimuth drive. CAST originally set an upper limit on the axion-to-photon coupling of $g_{a\gamma}~(95\%~$CL$) < 8.8\times 10^{-11}$ GeV$^{-1}$~\cite{Zioutas:2004hi,Andriamonje:2007ew} for axion masses up to 0.02~eV. Using the buffer gas technique~\cite{vanBibber:1988ge} the sensitivity was further extended to masses up to $\sim$1eV. The average upper limit provided by CAST with $^4$He (2005--2006) and $^3$He (2008--2011)is $g_{a\gamma}~(95\%~$CL$) \lesssim 2.3\times10^{-10}$ GeV$^{-1}$~\cite{Arik:2008mq,Aune:2011rx} and $g_{a\gamma}~(95\%~$CL$) \lesssim 3.3\times10^{-10}$ GeV$^{-1}$~\cite{Arik:2013nya}, respectively, with the exact value depending on the pressure setting.

It has been recently put forward~\cite{Irastorza:2011gs} that the helioscope technique can be substantially scaled up in size by building a large aperture superconducting magnet, and by extensive use of x-ray focusing optics and low background x-ray detection techniques. This has been the basis for the proposal of a new generation axion helioscope, the International Axion Observatory~\cite{Irastorza:1567109,Armengaud:2014gea}. IAXO envisions the construction of a large superconducting 8-coil, 20-m long toroidal magnet optimized for axion research~\cite{Shilon:2012te}. Between the coils, the IAXO magnet will host eight 60~cm-diameter bores, each of them equipped with x-ray optics~\cite{jakobsen2013} focusing the signal photons into $\sim$0.2 cm$^2$ spots that are imaged by ultra-low background Micromegas (MM) x-ray detectors~\cite{Aune:2013pna,Aune:2013nza}. The magnet will be built into a structure with elevation and azimuth drives that will allow solar tracking for $\sim$12 hours each day. In terms of signal-to-noise ratio, IAXO will be about 4$-$5 orders of magnitude more sensitive to Primakoff solar axions than CAST, which translates into a factor of $\sim$20 in terms of the axion-photon coupling constant $g_{a\gamma}$. That is, this instrument will reach the few $\times$10$^{-12}~{\rm GeV}^{-1}$ regime for a wide range of axion masses up to about 0.25~eV.

Although all the enabling technologies for IAXO exist, and the anticipated experimental parameters are considered reachable within the current state-of-the-art, the project is undergoing a stage of validation and prototyping as part of its technical design phase. We here present the design, construction and operation of a system that has been conceived as a technological pathfinder for IAXO, combining two of the techniques (optics and detector) proposed in the conceptual design of the project. The x-ray detector system is a shielded Micromegas (micromesh gas structure) detector made from radiopure materials and using the microbulk technique, an evolution of the detectors already used in the past in the CAST experiment. However, in the present configuration the detector is placed at the focal point of a $\sim$5 cm diameter, 1.3 m focal-length, cone-approximation Wolter I x-ray telescope comprised of thermally-formed (or ``slumped'') glass substrates deposited with multilayer coatings. This technology for x-ray optics is the one used in NASA's hard x-ray astrophysics NuSTAR satellite~\cite{harrison2013} and was identified in ~\cite{Irastorza:2011gs} to have the potential to cost-effectively cover the areas needed for IAXO with first-rate axion performance. Although CAST has already used  x-ray focusing optics for one of its four detector lines~\cite{Kuster:2007ue} (one of the spare optics from the ABRIXAS soft x-ray mission), this is the first time an x-ray optic is designed and built purposely for an axion application. It is also the first time a Micromegas detector is operated with an x-ray optic in an experimental setup. The system was installed in 2014 in one of the ports of the CAST magnet looking for axions during sunrise and is currently acquiring data in the experiment. The combination of the telescope and Micromegas detector provides the best signal-to-noise ratio obtained so far by any detection system of the CAST experiment and has a background rate of 5.4$\times$10$^{-3}\;$counts per hour in the signal spot area and in the energy region-of-interest (RoI), which is between 2-7~keV.

In Section \ref{sec:overview} we briefly describe the system in general, while the x-ray telescope and the x-ray detector are described in detail in Sections \ref{sec:optics} and 
\ref{sec:detector}, respectively. In Section \ref{sec:commissioning} the installation and commissioning of the system in CAST is presented, as well as its main performance parameters obtained. Finally, Section \ref{sec:conclusion} includes our conclusions and prospects.

\section{Overview of the system} \label{sec:overview}

\begin{figure}[htb!]
\centering
\includegraphics[width=140mm]{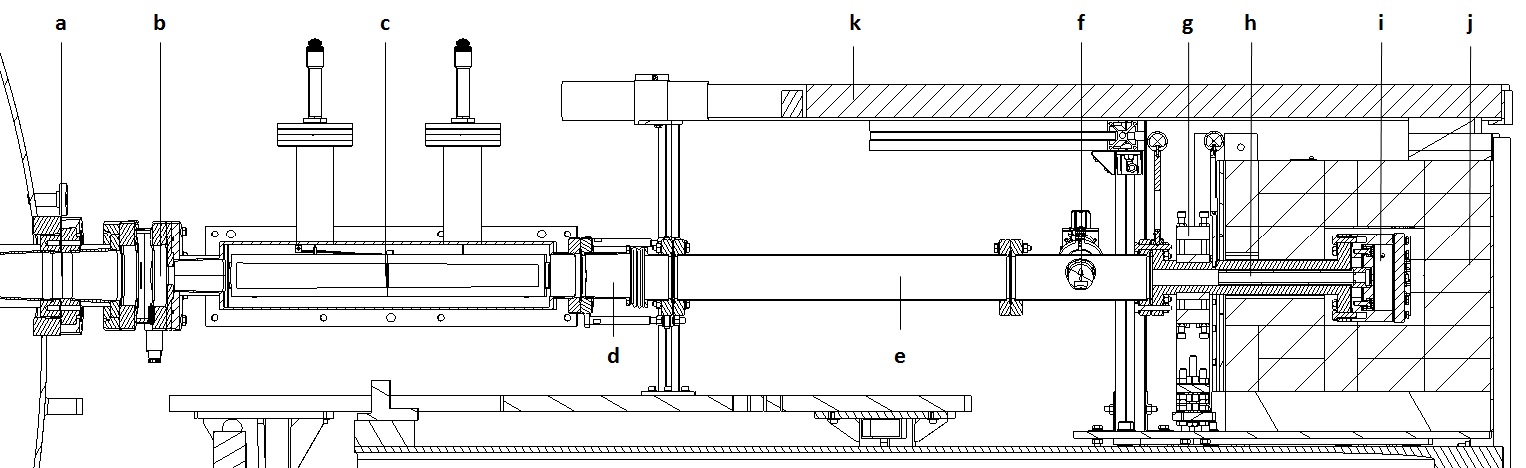}
\caption{Sketch of the new CAST sunrise detection line composed of a low background Micromegas detector placed at the focal point of an x-ray optic. The different parts of the line are described in detail in the text: gate valve (a), differential window (b), Wolter I x-ray optic (c), bellow (d), stainless steel interface tube (e), calibration system (f), precision stage (g), copper interface tube (h), Micromegas detector (i),
lead shielding (j) and muon veto (k).}
\label{fig:XRTMM}
\end{figure}

The new line is situated at one of the CAST sunrise magnet cold bores\footnote{At the CAST experiment, both ends of the magnet are equipped with x-ray detectors, three of which are Micromegas of the microbulk type. The magnet follows the Sun twice a day, during sunrise and during sunset. Depending on whether the detectors are taking data during the sunrise alignment or the sunset one, they are named sunrise Micromegas or sunset Micromegas respectively.}. It has been designed to fit the dedicated x-ray optic and the low background Micromegas detector system, and to align the two elements with the magnet bore. Both the optic and the detector have imposed some constraints on the line.
On one side, the detector must be placed at the focal point of the optic (already fixed because of space constraints of the experiment),
which fixed the length of vacuum elements. On the other side, the lead shielding must cover the detector tightly to effectively reduce the external gamma flux, which limits the range of the alignment elements.
A schematic view of the line is shown in Figure \ref{fig:XRTMM}.

Axion-converted x-rays coming from the cold bore would cross a gate valve (labelled {\bf a} in Figure \ref{fig:XRTMM}), which isolates the line from the magnet during commissioning periods, and a differential window ({\bf b}) made of 4~$\mu$m mylar, which protects the magnet cold bore vacuum in case of
a degradation of vacuum on the detector side. The optic case ({\bf c}) is directly bolted to this element ({\bf b}).
Between the Micromegas detector and the optic, there are four elements:
a bellow ({\bf d}), which allows to precisely align the detector with the optic
in combination with a precision stage ({\bf g}); a stainless steel tube ({\bf e}), whose length roughly fixes the detector position at the optic focal length; a calibration system ({\bf f}), composed of an actuator with an $^{55}$Fe source and a copper interface tube ({\bf h}), which is screwed to the detector chamber ({\bf i}) and forms also part of the shielding. An octagonal Polytetrafluoroethylene (PFTE) cassette covers the inner tube's walls to block the fluorescence coming from the copper, which might be activated by external radiation. 
As shielding, there is a lead layer of 10~cm thickness ({\bf j}), which reduces the external gamma flux, and a plastic scintillator coupled to a photomultiplier ({\bf k}), which works as an active muon veto. The optic and the Micromegas detector are described in detail in the following.

\section{Optics} \label{sec:optics}

\subsection{Overview} \label{subsec:opOver}

The x-ray optic designed for use with Micromegas detectors was made following the same techniques \cite{caltechauthors24759, koglin2009, craig2011, hailey2010} developed for NASA's NuSTAR satellite mission~\cite{nustar2013}. In this approach, flat-panel glass is thermally formed (slumped) in cylindrical shapes and then cut into conical (truncated) cones. After cleaning, the glass substrates are deposited with multilayer coatings to enhance x-ray reflectivity. These individual x-ray reflectors are then epoxied into a precision assembly that results in a reflective x-ray optic that approximates a Wolter I geometry~\cite{Petre85}.

\begin{table}[!htb]
\begin{center}
\begin{tabular}{ll}
Property	&  Value\\
\hline
\hline\\[-3mm]
Mirror substrates				& glass, Schott D263	\\
Substrate thickness				& 0.21~mm      \\
$L$, length of upper and lower mirrors		& 225~mm      \\
Overall telescope length			& 454~mm      \\
$f$, focal length				& 1500~mm      \\
Layers						& 13      \\
Total number of individual mirrors in optic	& 26      \\
$\rho_{\rm{max}}$, range of maximumm radii	& 63.24--102.4~mm      \\
$\rho_{\rm{mid}}$, range of mid-point radii	& 62.07--100.5~mm     \\
$\rho_{\rm{min}}$, range of minimum radii	& 53.85--87.18~mm      \\
$\alpha$, range of graze angles			& 0.592--0.968 degrees      \\
Azimuthal extent				& Approximately 30 degrees      \\
\end{tabular}
\end{center}
\caption{Properties of the segmented-glass telescope at CAST.}
\label{tab:properties}
\end{table}  

For solar axion experiments, x-ray telescopes made from segmented glass substrates offer two distinct advantages, compared to integral-shell telescopes made via replication or from large glass/ceramic blanks.  First, unusual, azimuthally asymmetric designs can be fabricated to minimize the amount of space required for vacuum lines and supporting hardware.  This consideration is particularly important for CAST, since during its lifetime, the experiment has undergone several upgrades that have required the addition of new structures.  As a result, there is limited space for new instruments, so the use of segmented-glass fabrication technique was critical in building just enough of an x-ray optic to cover a single magnet bore that has a diameter of 43 mm.  Second, one can deposit multilayer coatings optimized to maximize the x-ray reflectivity exactly at the peak of the axion spectrum, thus increasing the signal-to-noise and sensitivity. This unique advantage is discussed in more detail below.

\subsection{Design of the optic} \label{subsec:opDesign}

\begin{figure}[htb!]
\centering
\includegraphics[width=115mm]{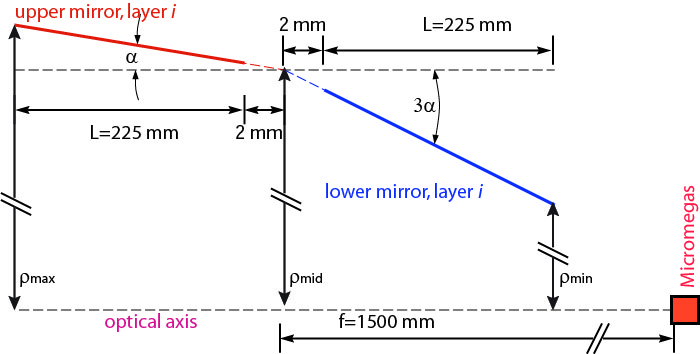}
\includegraphics[width=90mm]{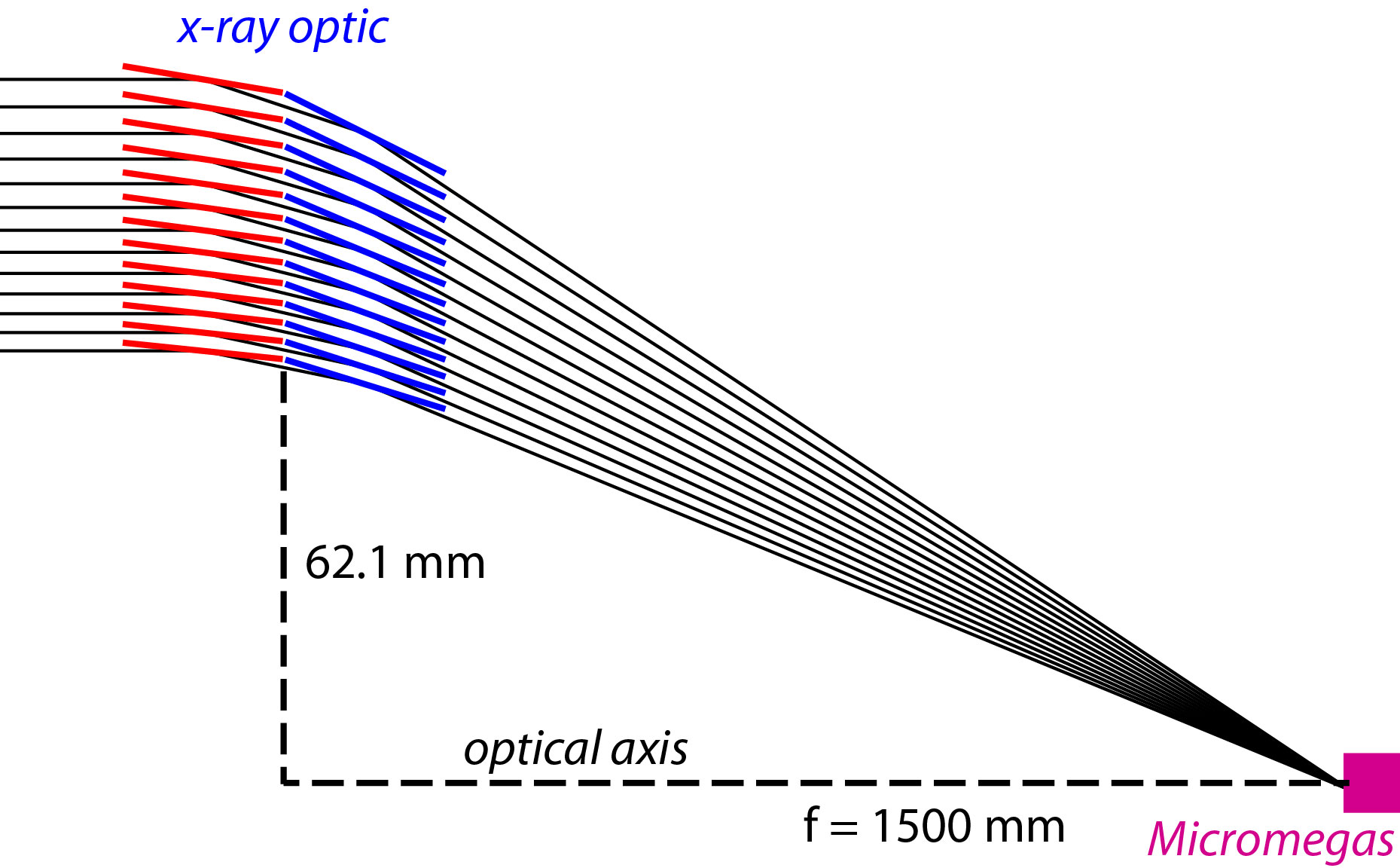}
\caption{Top: Schematic cross-section drawing of the CAST segmented-glass telescope.  A single layer i of the telescope is shown, with key dimensions described. Bottom: A 10:1 stretched ray-trace of the optic, where the vertical dimension has been multiplied by a factor of 10 to better show the relative position of the individual mirrors with respect to each other. Each black line represents a single photon, coming from infinity, incident in the middle of each of the 13 nested shells. The cone-approximation Wolter I design focuses all the light to a small region on the Micromegas detector.}
\label{fig:optics}
\end{figure}

Since the angular size and spectrum of potential solar axion emission is known, a figure of merit can be defined that can be used to optimize an x-ray telescope for an axion helioscope~\cite{ Armengaud:2014gea,Irastorza:2011gs}. For CAST, the optical design space is quite limited for a number of reasons.  First, the new x-ray optic, Mircomegas detector, vacuum vessels and shielding must conform to the limited physical space of the new beamline.  This limited the optic to a focal length of 1.5 meters.  Another important constraint was the re-use of existing infrastructure originally fabricated for the NuSTAR x-ray telescope \cite{craig2011,hailey2010}. This set a minimum radius and defined the length of upper (parabolic-like) and lower (hyperbolic-like) shells.  Based on these two constraints, a cone-approximation design was generated, compatible with a Micromegas serving as the focal plane and compliant with overall CAST constraints. Table~\ref{tab:properties} lists the properties of the CAST telescope, Figure \ref{fig:optics} at the top shows a two-dimensional cross-sectional schematic of the optic and Figure \ref{fig:optics} bottom shows a 10:1 stretched ray-trace of the optic.

The next step was to design the x-ray-reflective multilayer coating, using the general approach discussed in detail in \cite{jakobsen2013}. For this particular optic for CAST, given the limited options in changing the key parameters for the optical design, we fixed the multilayer material system to be Pt/C and decided to limit the different multilayer recipes to four.   A full explanation of the optimization process is given in~\cite{JakobsenPhD}. Table~\ref{tab:multilayer} presents the key properties of each of the four multilayer recipes used for this optic.

\begin{table}[!htb]
\begin{center}
\begin{tabular}{cccccc}
Design	&  Applied to layers & $N$ & $d_{\rm{min}}$  & $d_{\rm{max}}$  & $\Gamma$ \\
\hline
\hline\\[-3mm]
1	& 1--3		&   2	&	11.2& 22.5	& 0.46	\\           
2	& 4--6 		&   3	&   6.9	& 18.5	&0.49	\\
3	& 7--10		&   4	&   5.2	& 16.2	&0.41	\\	
4	& 11--12	&   5	& 	4.9 & 13.6	& 0.40	\\		
\end{tabular}
\end{center}
\caption{Multilayer coating prescriptions of the Pt/C multilayers.}
\label{tab:multilayer}
\end{table}  

\subsection{Expected CAST performance} \label{sec:performance}

The point spread function (PSF) and effective area EA (throughput) of the x-ray optic will be calibrated in detail after the scientific campaign.  In the meantime, we have estimated the performance of the optic, as part of the CAST helioscope, based on measurements of the witness samples from the multilayer depositions (for details see \cite{JakobsenPhD}) and the in situ measurements with a Cool-X source \cite{coolx} (see Section \ref{sec:commissioning}).

The assumption that the optic has the same PSF as the NuSTAR telescopes~\cite{nustar2013} is supported by the Cool-X data.  Specifically, a baseline model has been assumed where the figure errors produce a PSF well-described by a narrow Gaussian core of 30 arcsec and a half-power diameter (HPD, sometimes also referred to as the 50\% encircled energy function [EEF]) of 58 arcsec.

As a detailed calibration of the new XRT has not yet been performed, to be conservative, for the purpose of defining the extraction region for the CAST science analysis,  a PSF that is 1.3 times worse than expected is assumed.   Although this will result in a slightly higher estimate of the background, this conservative step minimizes the risk of missing a putative signal.  After the detailed calibration is performed, the analysis can be repeated with smaller extraction regions.

To simulate the expected signal spot size, the emission region was modelled as a 3 arcmin diameter, located 1 A.U. from the CAST experiment, and assumed a solar axion differential spectrum given in \cite{Andriamonje:2007ew} of the form

\begin{equation}
\frac{d\Phi}{dE} \propto \frac{E^{2.481}}{\exp{(E/1.205)}} \; ,
\end{equation}

\noindent where $E$ the axion energy in keV. 
 
\begin{figure}[htb!]
\centering
\includegraphics[height=0.35\textwidth]{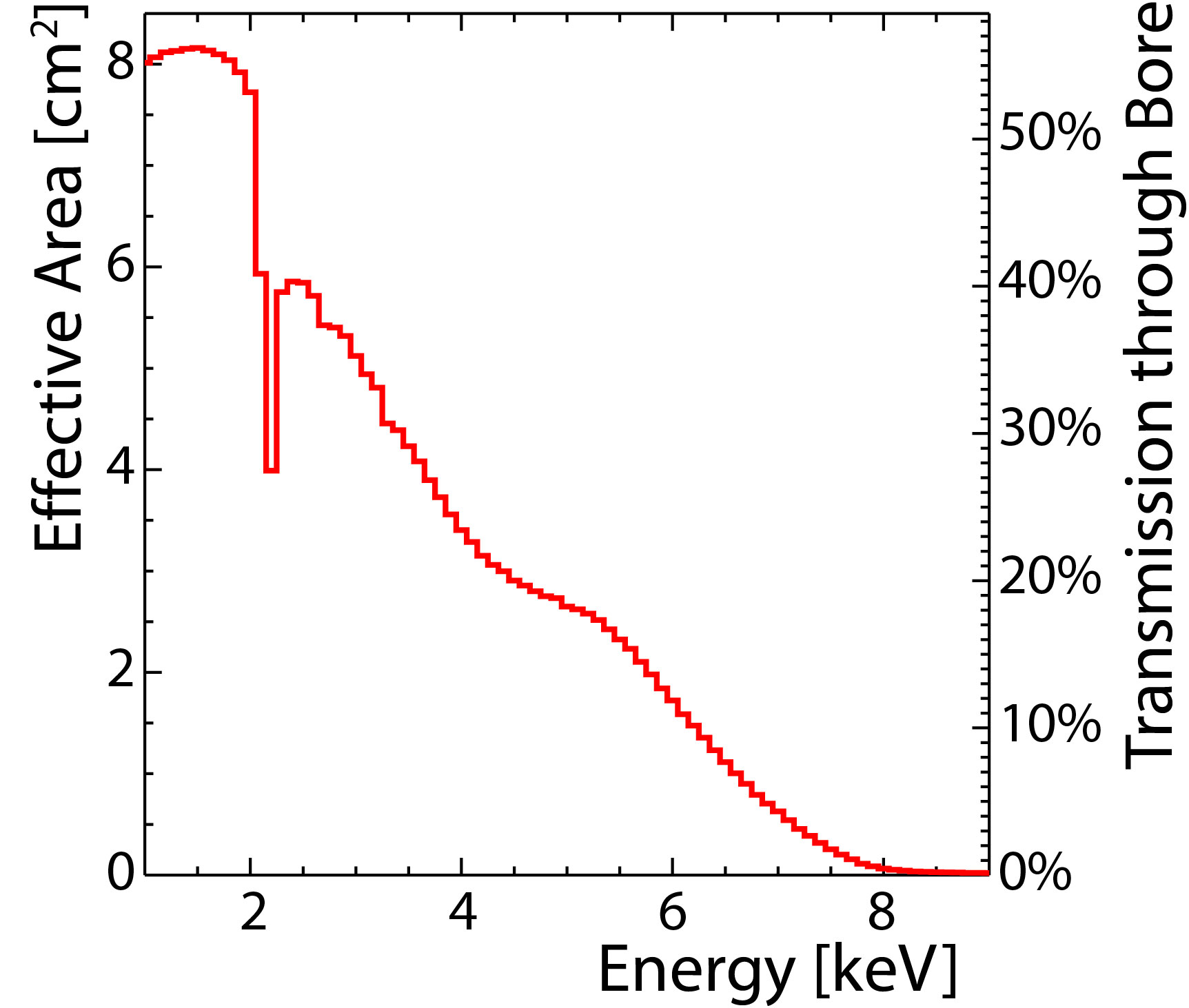}
\caption{Effective area (cm$^2$) and transmission (percentage) versus energy.  The transmission is computed by dividing the effective area (EA) by the geometric area of the 43~mm diameter bore. The EA was modelled using the as-built optical prescription, assuming a half-power diameter of 75 arcsec and that the solar axion emission comes from a uniformly distributed 3 arcmin disc.}
\label{fig:throughput}
\end{figure}
 
Millions of photons were ray-traced through the optic, before they were registered on an infinitely large sheet at the focal plane to calculate the telescope throughput (effective area). Figure~\ref{fig:throughput}, plots the EA versus energy. It should be noted that this EA is lower than what would be computed using standard practice in the x-ray astronomy community, which reports EA for a point-source on-axis. This approach directly accounts for the loss of throughput and vignetting when imaging an extended source. Finally, a conservative lower energy limit of 1~keV has been set, since no previous calibration of a segmented glass x-ray telescope in the soft x-ray regime exists.

Because the bore only illuminates approximately a 30$^{\circ}$ azimuth ``wedge'' of a surface of revolution, the resultant focused light is not circularly symmetric.  For a point-source, the resultant image would resembled a bow-tie.  For a source with a 3 arcmin extent, the resultant focused spot is roughly rectangular. Rather than circular extraction regions (e.g., the commonly used half-power diameter), rectangular extraction regions are used.  As the extraction regions increase in area, more of the photons will be captured.  Table~\ref{tab:flux} presents the area of rectangular extraction regions required to enclose increasingly larger fractions of the total reflected signal.

\begin{table}[!ht]
\begin{center}
\begin{tabular}{cccccc}
Fraction of the flux captured	& Geometric area   \\
(\%)	& (mm$^2$)   \\
\hline
\hline\\[-3mm]
47.3  &  1.44	\\	
72.0  &  3.08	\\
90.6  &  5.76	\\
99.5  &  13.6	\\	
\end{tabular}
\end{center}
\caption{Extraction region sizes and captured flux.}
\label{tab:flux}
\end{table}

\section{Detector} \label{sec:detector}

\begin{figure}[htb!]
\centering
\includegraphics[width=120mm]{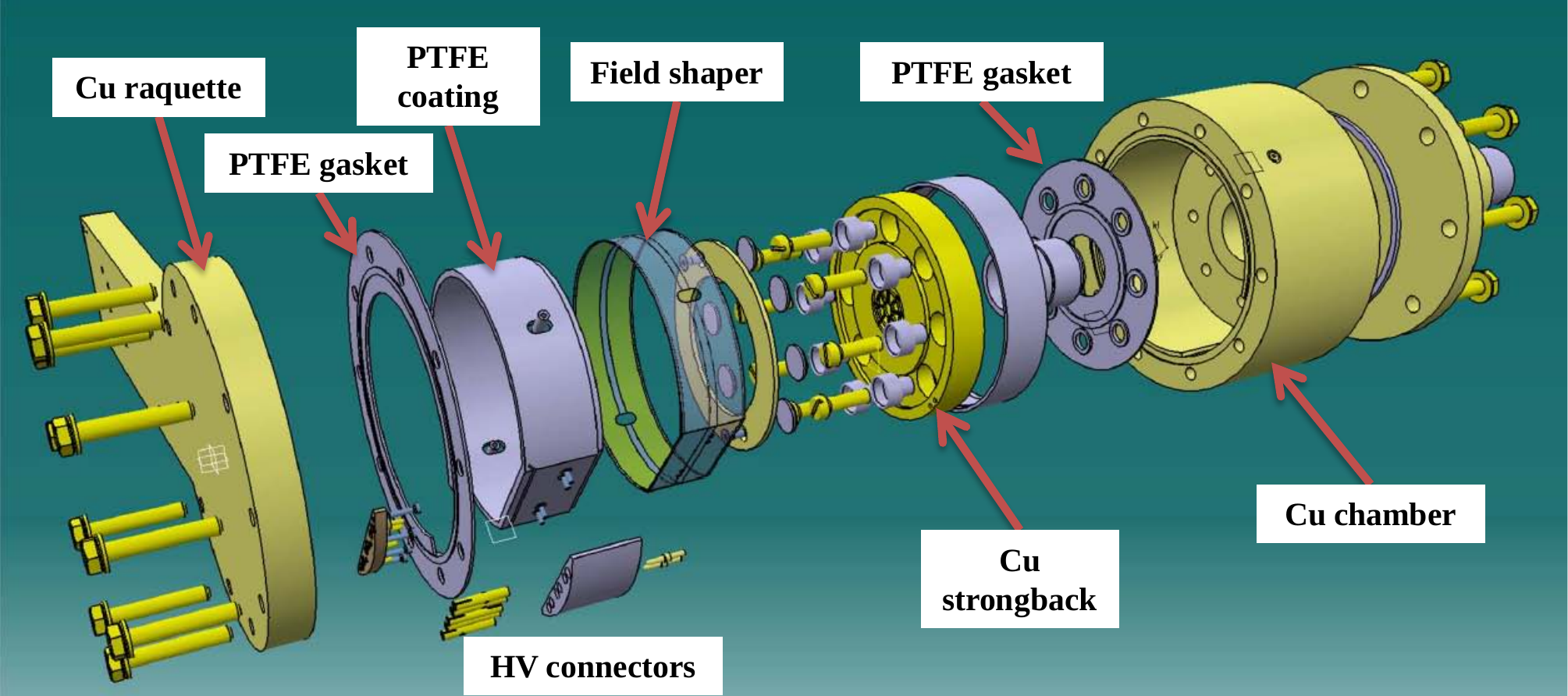}
\caption{Exploded view of the sunrise Micromegas detector, where the different parts of the chamber are labelled. The components are described in detail in the text.}
\label{fig:SRmMDesign}
\end{figure}

\begin{figure}[htb!]
\centering
\includegraphics[height=0.25\textwidth]{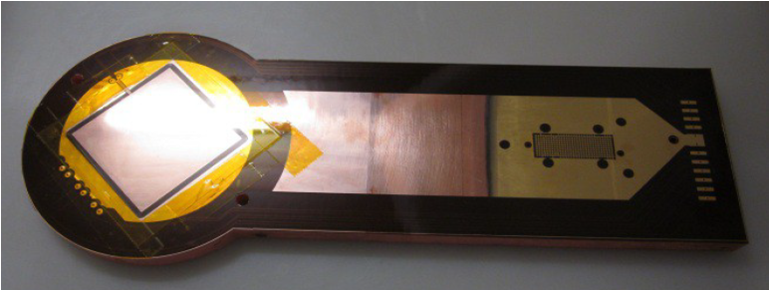}
\caption{A photograph of the new sunrise Micromegas readout glued on the copper support. At the left of the image is the squared active area, while the strip signals are extracted from a central connector situated on the right part. The high voltage is fed through squared pads situated on the right part and reach the mesh and the field shaper through lines situated at the sides of the printed board.}
\label{fig:SRmMDetector}
\end{figure}

The design of the new detector (Figure \ref{fig:SRmMDesign}) is the prime example of the current state-of-the-art in low background techniques \cite{Aune:2013pna}, developed for the Micromegas detectors. In contrast to previous designs \cite{Abbon:2007pa} the body and the chamber of the detector are made of $\sim$18~mm thick radiopure copper which has been carefully cleaned; moreover, all the gaskets consist of radiopure PFTE. A new field shaper has also been installed, printed on a kapton circuit and integrated in the chamber. The field shaper, which increases the uniformity of the drift field and reduces border effects, is externally covered by a 3~mm thick PFTE coating in order to block copper fluorescence. In addition, the high voltage (HV) connections were implemented in the detector's printed board, allowing for an easy extraction of signal and voltages out of the shielding. A picture of the new Micromegas readout plane is shown in Figure \ref{fig:SRmMDetector}.

\begin{figure}[htb]
\centering
\includegraphics[height=0.33\textwidth]{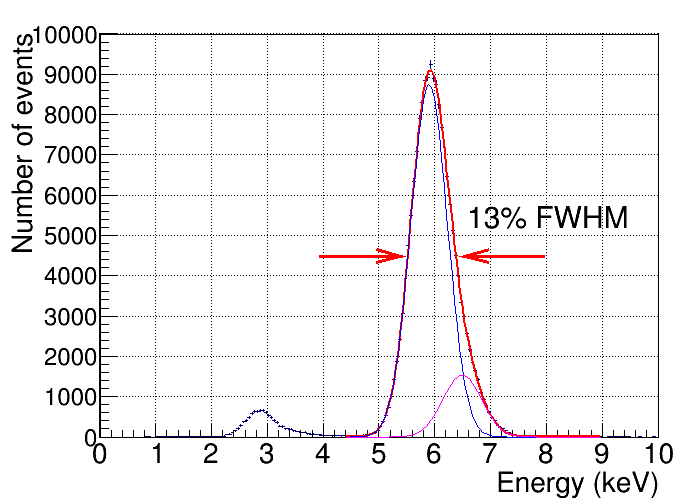}
\includegraphics[height=0.34\textwidth]{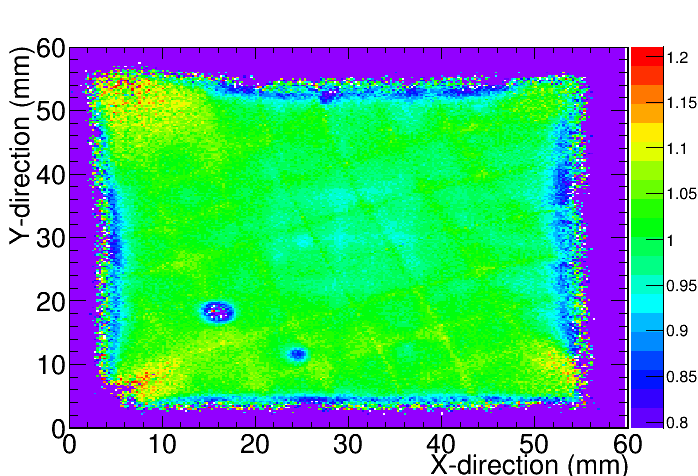}
\caption{Left: $^{55}$Fe calibration spectrum of the sunrise Micromegas detector. The main peak has been fitted to two gaussian functions (blue and magenta lines, in red the sum of them), corresponding to the K$_\alpha$ (5.9~keV) and K$_\beta$ lines (6.4~keV). Right: Gain uniformity of the sunrise Micromegas detector. The dead areas (in purple) show lower values than unity (in green) and lie outside the axion signal area.}
\label{fig:SRTelesPerformance}
\end{figure}

New Micromegas planes of the microbulk type have been manufactured for the new line.  The readout has been modified with respect to previous versions \cite{Andriamonje:2010zz}: the strip pattern has a smaller pitch (500~$\mu$m instead of 550~$\mu$m) while keeping the same active area ($60~\times~60$~mm$^2$), which has increased the number of strips to 120 per axis (as opposed to 109). This new design is the result of studies made on the Micromegas detectors in order to improve their performance and of the enhancements of the manufacturing technique \cite{Aune:2013pna}. So far, it is the Micromegas with the best performance at CAST with a 13$\%$ (FWHM) at 5.9~keV (Figure \ref{fig:SRTelesPerformance}, left), close to the 11\%~(FWHM) reached by small non-pixelated microbulk prototypes (3.5~cm diameter) \cite{Iguaz:2012fi}. The detector shows a good homogeneity of the gain in the active area (Figure \ref{fig:SRTelesPerformance}, right). 

\begin{figure}[htb!]
\centering
\includegraphics[width=140mm]{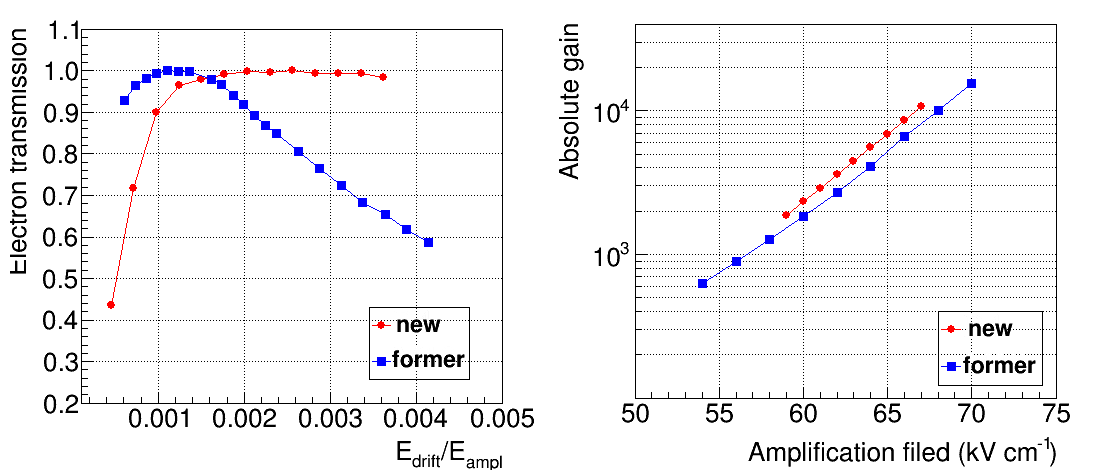}
\caption{Dependence of the electron transmission on the ratio of drift and amplification fields (left) and of the absolute gain on the amplification field (right) for a new CAST sunrise Micromegas detector (circled red line) and a former design (squared blue line), characterized in Ar+2\%iC$_4$H$_{10}$ at 1.4 bar.}
\label{fig:SRTelesPerformance2}
\end{figure}

This improvement in performance is due both to a better manufacturing technique and to the implementation of an optimal ground strategy, described at the end of the section.
The electron transmission curve of the detector, which indicates the percentage of primary electrons that will pass from the conversion to the amplification area of the detector through the micromesh, depends on the ratio of the fields in the two regions and defines the range of drift fields for an optimum detector performance. In the new detector this range has been enlarged with respect to the previous ones (Figure \ref{fig:SRTelesPerformance2}, left), due to a change of the geometry of the micromesh holes (diameter and pitch) and the increased uniformity provided by the field shaper. The gain curve (Figure \ref{fig:SRTelesPerformance2}, right), mainly defined by the amplification gap and the gas conditions, is similar to previous designs.

\begin{figure}[htb!]
\centering
\includegraphics[height=47mm, width=47mm]{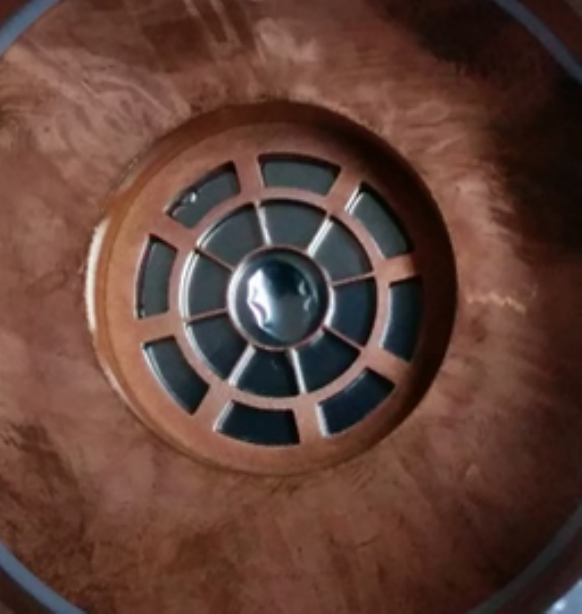}
\includegraphics[height=52mm]{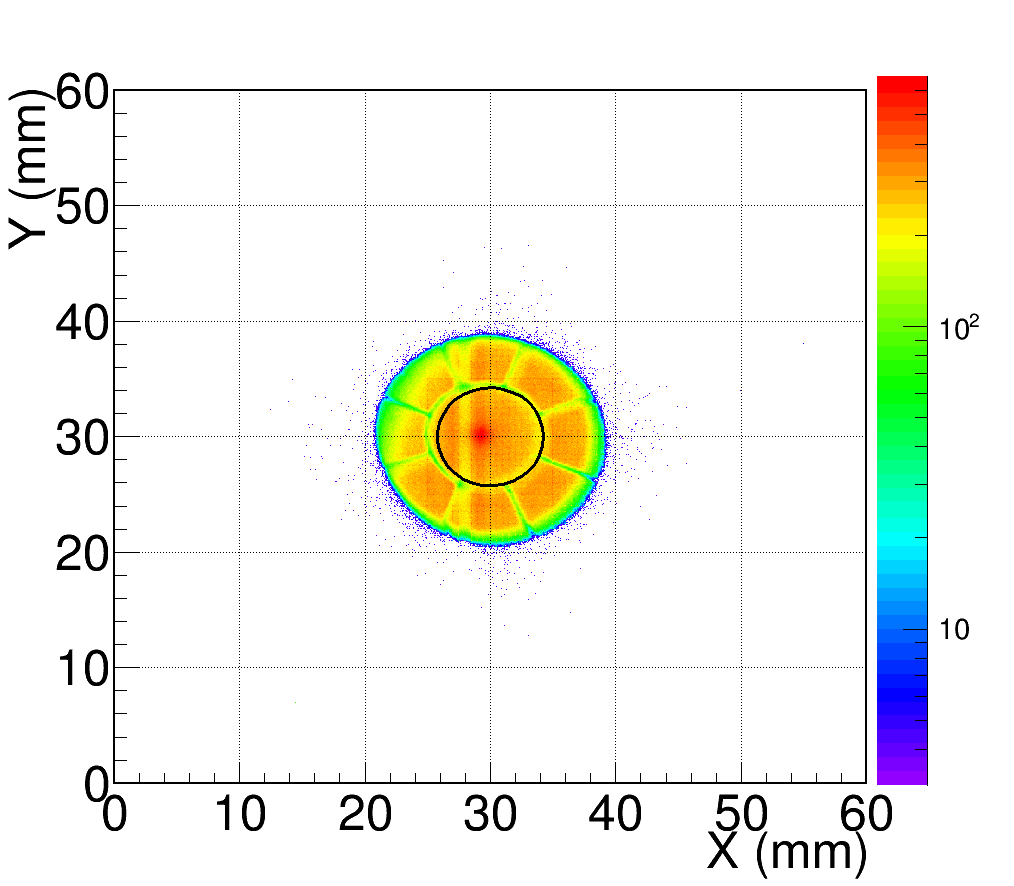}
\caption{Left: Photo of the new copper strongback with the spider web design.
Right: Intensity plot with the projection of the x-ray window (cathode) in the Micromegas detector during the calibrations.}
\label{fig:SRTelesStrongBack}
\end{figure}

After focusing by the optic, the signal region is expected to be reduced to a $<$5~mm spot, therefore the x-ray window (cathode) strongback pattern has been modified to resemble a spider-web design with a central hole of 8.5~mm in diameter, big enough to contain the expected axion signal image. The focused x-rays go through the 4~$\mu$m-thick aluminized polypropylene window avoiding the grid structure, which was responsible of a $\sim$10$\%$ of efficiency loss in previous setups. Consequently, the efficiency of the new detector is increased. A photo of the new strongback design together with its projection during calibrations in the detector are shown in Figure \ref{fig:SRTelesStrongBack}.

Signals induced both in the mesh and in the strips come out from the active area by printed paths situated in the medium layer of the Micromegas detector. The mesh pulse is subsequently amplified by a CANBERRA 2014 preamplifier and an ORTEC 474 Timing amplifier and then duplicated. One of these copies is then sampled at 1 GHz frequency in a 2.5~${\rm \mu}$s window and recorded by a 12-bit dynamic range VME digitizing board, MATACQ (MATrix for ACQuisition) \cite{Breton:2005db}.

Meanwhile, the strip pulses pass to an interface card, where they are distributed to four flexible cables connected to an AFTER (ASIC For TPC Electronics Readout)-based Front-End Card board \cite{Baron:2008zza, Baron:2010pb}. Each one of the four AFTER ASICs collects and samples the strip signals continuously at 100 MHz in 512 samples per channel, recording a window of $\sim$5${\rm \mu}$s, which is longer than the maximum drift time of charges created in the active volume. The readout electronics are triggered by the second copy of the mesh pulse. Then, the analogue data from all channels is digitized by an Analog-to-Digital Converter (ADC). Finally, a pure digital electronics card, the FEMINOS board \cite{Calvet:2014zva}, gathers ADC data, performs the pedestal subtraction and sends them to the data acquisition (DAQ) system by means of an ethernet connection.

Special attention has been given to grounding in the electronics design: signal paths are surrounded by a ground layer at the detector, the interface card and the flat cables to avoid any coupling; the AFTER-based cards and the preamplifier are fixed to the inner part of a Faraday cage to minimize induced noises and each of the high voltage lines has a dedicated low-frequency filter to dim ripples from HV sources.

\begin{figure}[htb!]
\centering
\includegraphics[height=0.29\textwidth]{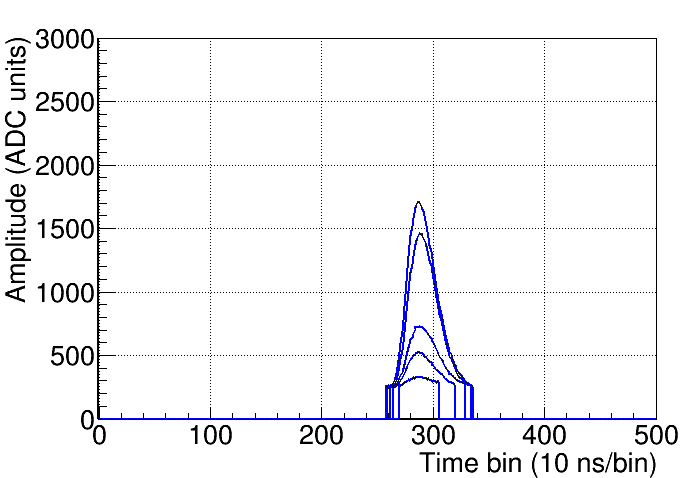}
\includegraphics[height=0.29\textwidth]{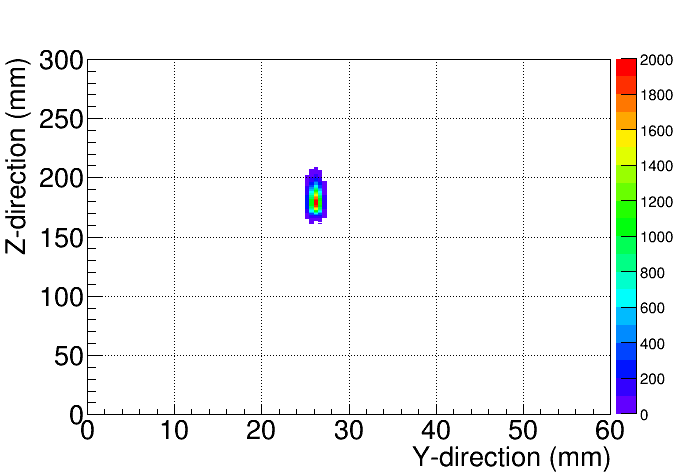}
\includegraphics[height=0.29\textwidth]{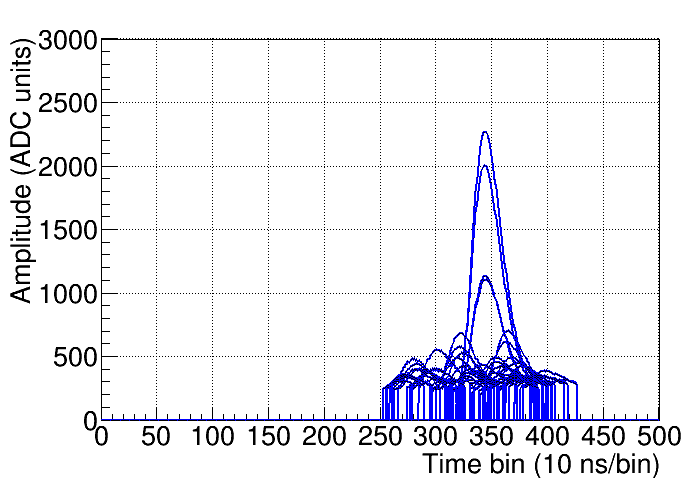}
\includegraphics[height=0.29\textwidth]{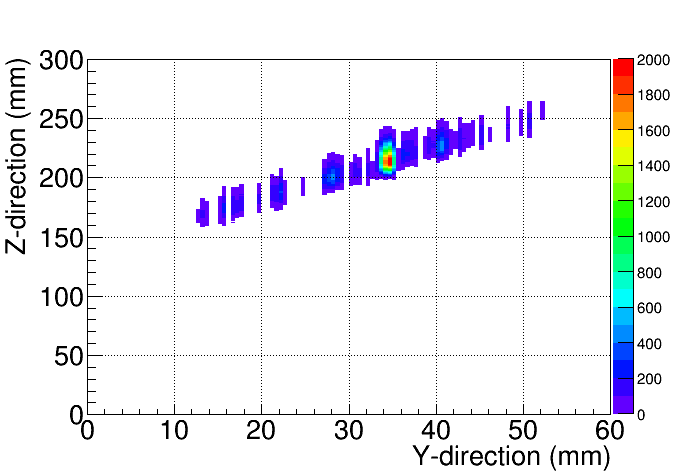}
\caption{Left: y-strip pulses (recorded in zero-suppression mode) of an x-ray (top) and a background event (bottom) in a CAST Micromegas detector. Right: The yz view of the same events.}
\label{fig:3DEvent}
\end{figure}

The 3D reconstruction of an event is made combining the strip pulses, whose temporal position determines the z axis, and the detector decoding, used for the spatial coordinates, x and y on the detector plane. The charge collection of each event is projected in both spatial and temporal direction. As an example, the signals induced on strips and acquired by the AFTER-based electronics and the resulting \textit{yz} projection is shown in Figure \ref{fig:3DEvent}.

\section{Installation and commissioning in CAST} \label{sec:commissioning}

\begin{figure}[htb!]
\centering
\includegraphics[height=55mm]{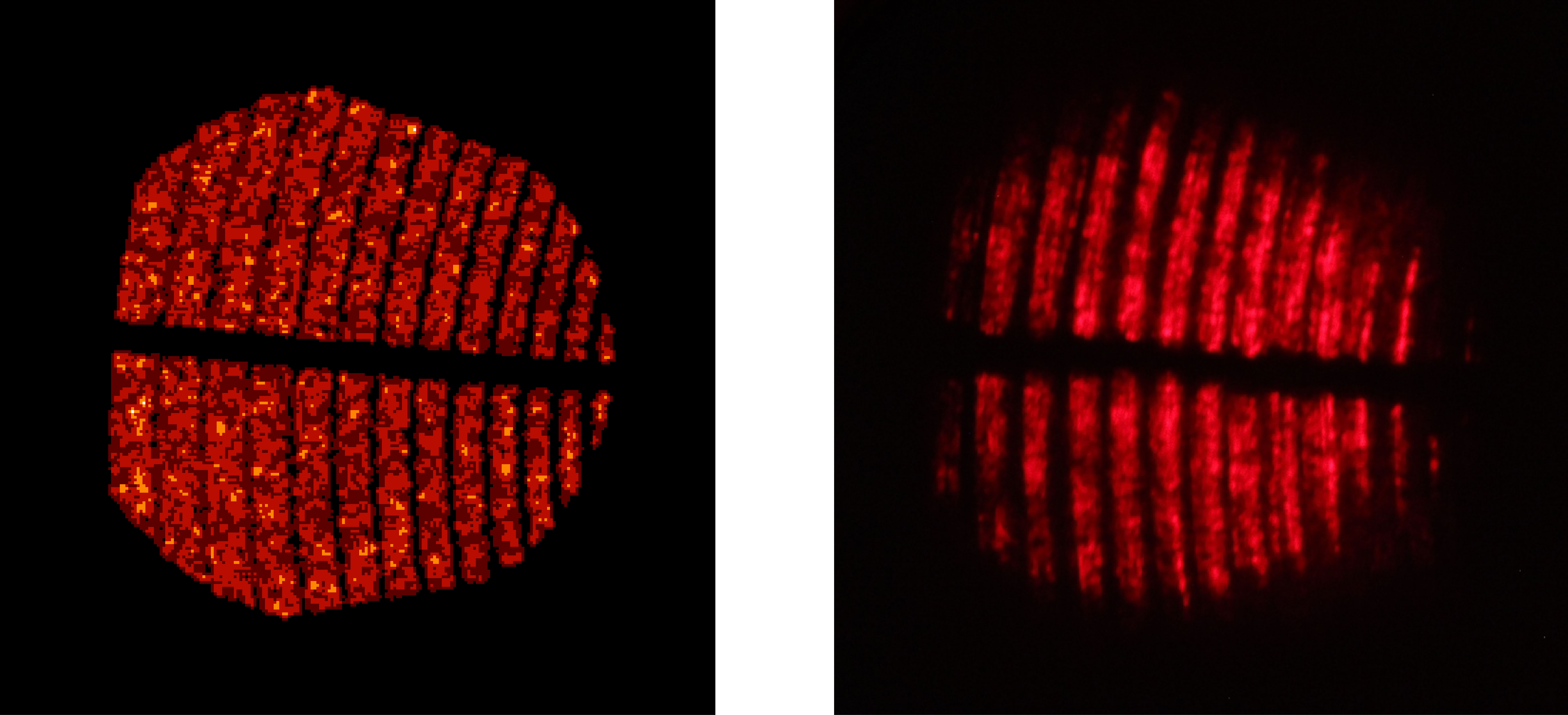}
\caption{Illumination of the X-ray optic with a HeNe (visible wavelength) laser, after installation and alignment of the telescope into CAST. These images represent the light reflecting and diffracting through the optic, just after the exit aperture. The large gap is due to obscuration by the graphite spacers used to fabricate the optic. Left:  Simulation based on ray-trace techniques and ad hoc corrections to account for visible wavelength diffraction. Right: Photograph of the HeNe laser reflection and diffraction through the optic.}
\label{fig:laser}
\end{figure}

The new SRMM detector system was installed at CAST in August 2014. 
Rough alignment of the optic and its housing was performed using standard surveying techniques.  The next step was to locate a HeNe laser on the bore of the magnet, and using fiducials built into the support structure of the optic, to do a fine alignment.   The final step was to insert a diffuser into the visible laser beam, to illuminate the entrance bore of the x-ray telescope.   Because a Wolter design is reflective, the optic focuses visible wavelength well (although there is some diffraction), and we compared the measured HeNe transmission at the exit aperture of the x-ray telescope with our ray-trace results.  The measurement and simulation are show in Figure~\ref{fig:laser}, and the agreement between the two indicated that the telescope was aligned to better than 30 arcsec.

\begin{figure}[htb!]
\centering
\includegraphics[height=0.25\textwidth]{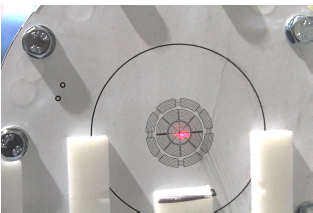}
\includegraphics[height=0.25\textwidth]{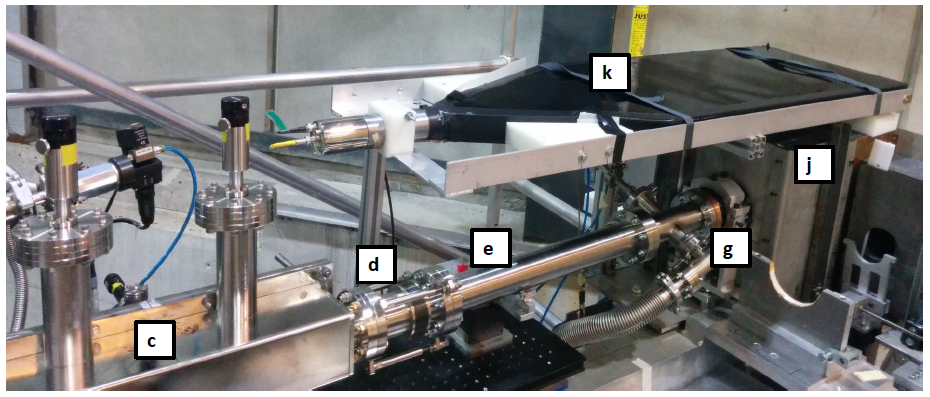}
\caption{Left: The flange used to align the detector: the laser is focused at the center of the designed pattern, as expected. Right: A picture of the line installed on the CAST platform. The annotated elements follow the description of Figure 1.}
\label{fig:photoLine}
\end{figure}

After the telescope was aligned using the theodolite, the vacuum components for the differential pumping \cite{Abbon:2007pa} were installed on either side (elements b, d, e and h of Figure \ref{fig:XRTMM}). The alignment of the Micromegas detector followed, in such a way that the spot be focused at the center of the sensitive area of the detector. In order to accurately fix the position of the detector, a polyethylene support was built. An XY stage was designed as a tailor-made solution to cope with the total weight of the  full assembly ($\sim$30~kg) and the moment distribution fixed by geometrical constraints given by the shielding volume and the available room on the detector platform. The alignment was performed with a blank flange at the place of the detector cathode; the flange had a design of the detector's cathode hole-pattern imprinted, on which the parallel laser beam was focused (Figure \ref{fig:photoLine}, left). When the chamber position was defined, the bolted unions of the XY stage were locked, fixing thus the position of all line components. Then, the detector replaced the alignment flange and the chamber support was dismounted.

The link between the vacuum line and the detector is made by a 20~mm-thick copper-pipe interface with a PFTE coating. The aperture of the pipe has a diameter of 25~mm. A shielding of 10$\;$cm of lead was then built around the detector, leaving only one side at 7$\;$cm because of space constraints. The inner shielding is the detector chamber itself, made of 18~mm radiopure copper. Following the results of extended studies performed \cite{Aune:2013pna}, a plastic scintillator for the detection of cosmic muons was installed on the top of the shielding. The scintillator covers a part of the pipe that connects the Micromegas with the optic, as external radiation interacting with it is thought to be the source of a significant amount of the background \cite{Aune:2013pna}. With the information from the scintillator, the induced x-ray-like events can be discriminated by the off-line analysis. A picture of the installed line can be seen in Figure \ref{fig:photoLine}, right.

\subsection{Stability of the spot position}

\begin{figure}[htb!]
\centering
\includegraphics[width=0.30\textwidth]{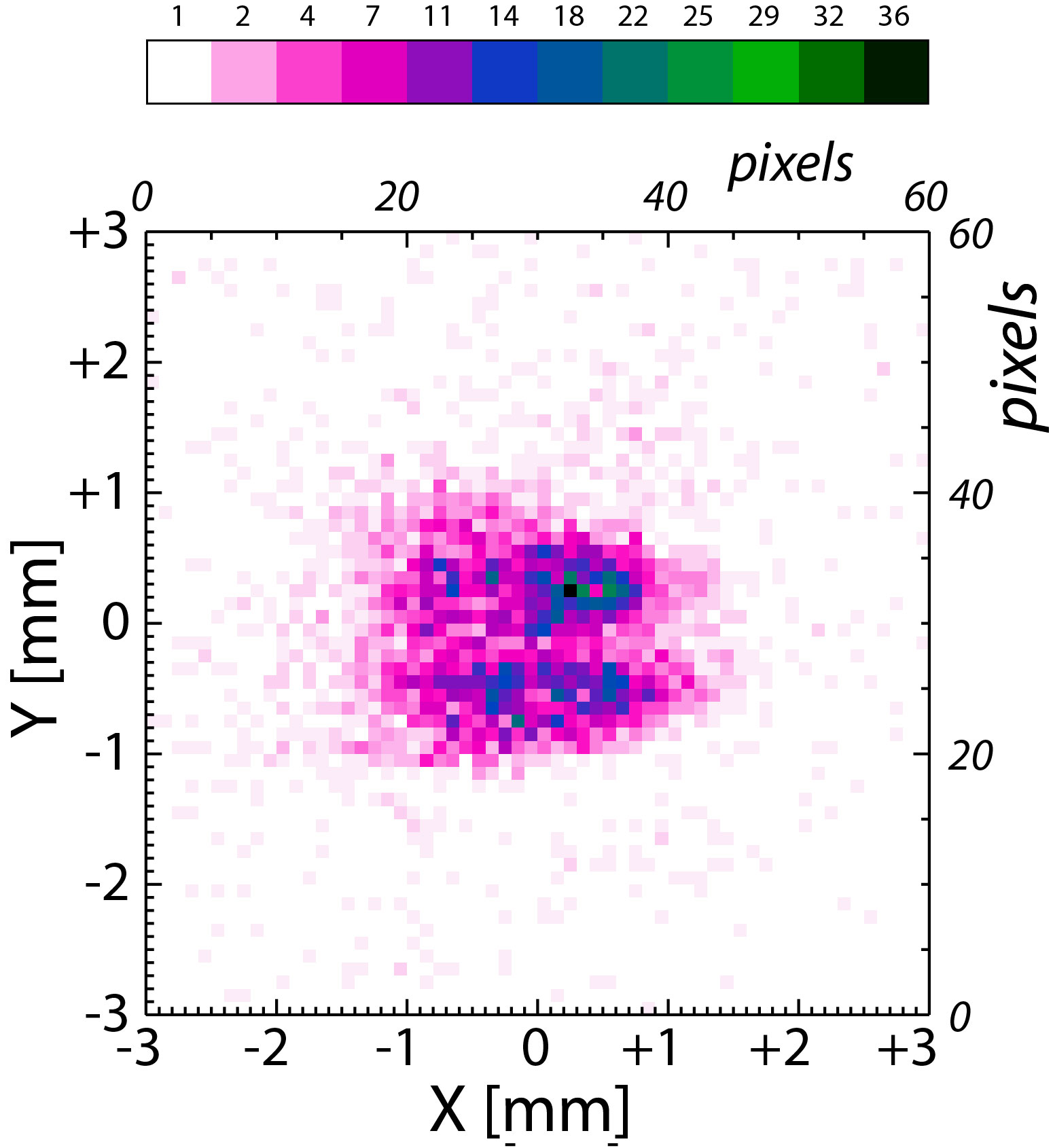}
\includegraphics[width=0.30\textwidth]{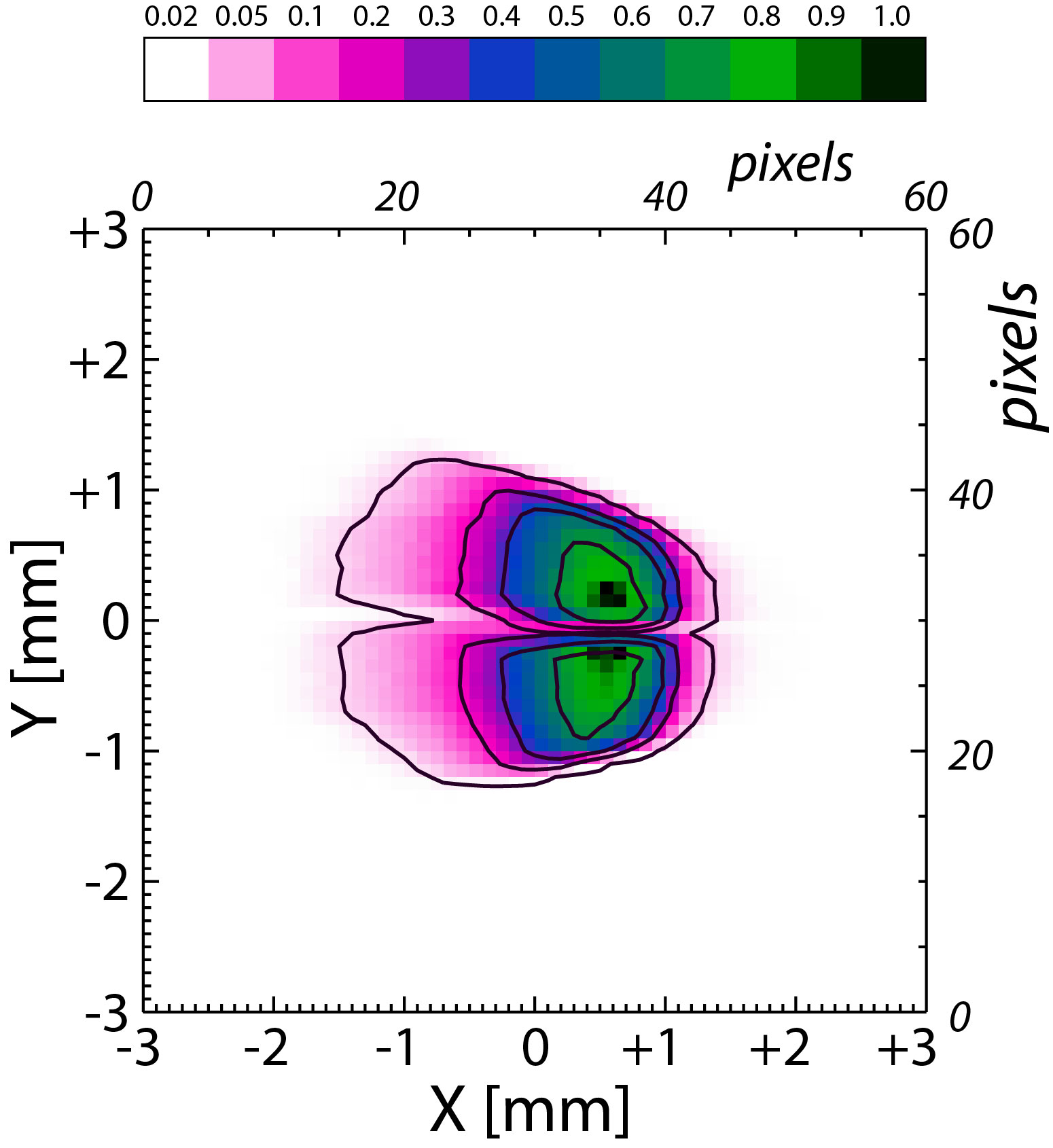}
\includegraphics[width=0.30\textwidth]{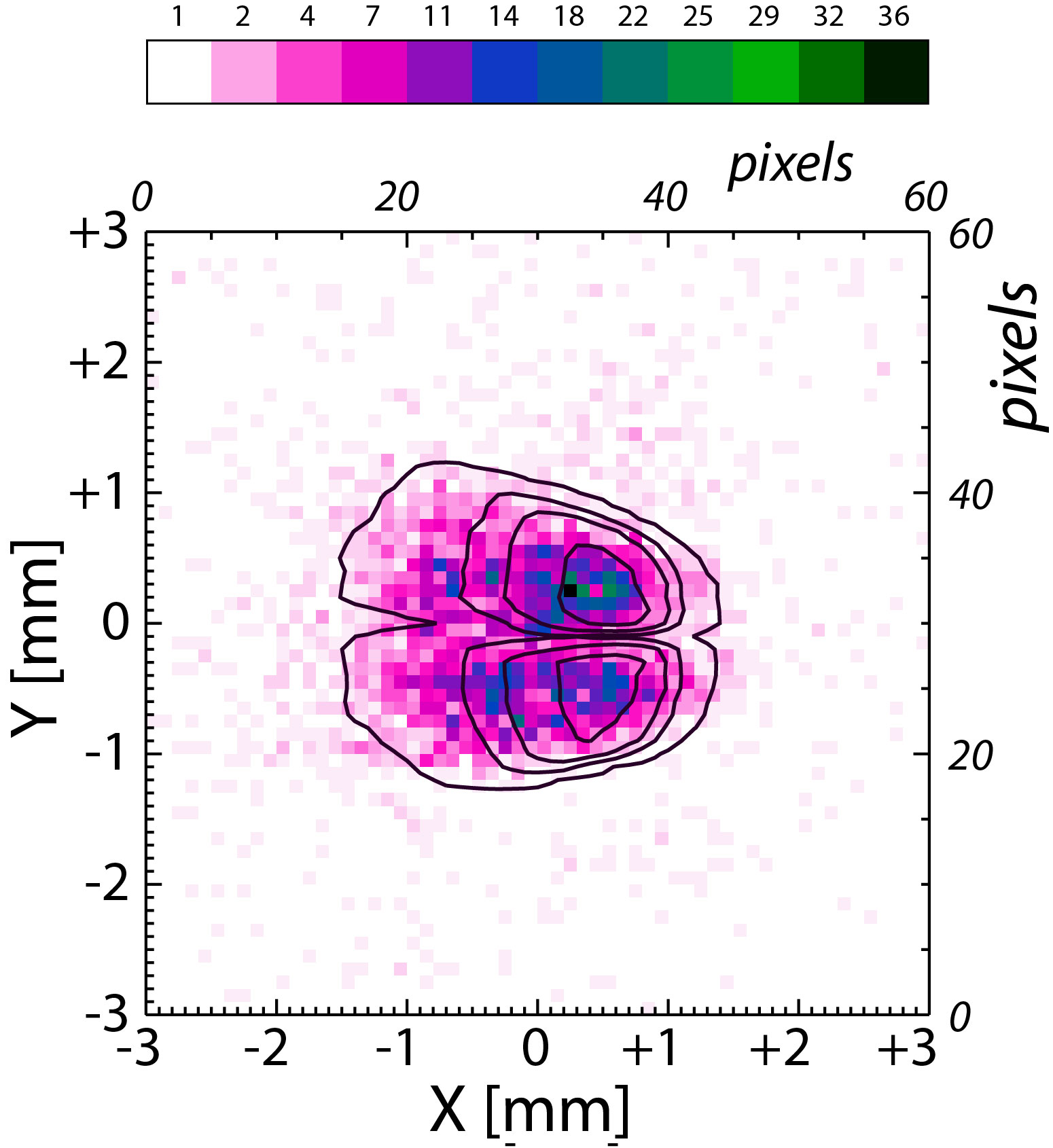}
\caption{Cool-X x-ray source illumination of the new x-ray telescope. Left: image measured by the Micromegas detector. Middle: simulation, assuming the Cool-X x-ray emission comes from a uniform 6~mm diameter spot. Contour levels are 6\%, 30\%, 50\% and 80\% of maximum intensity. Right: data of the Micromegas detector, now with the simulation contours over-plotted.}
\label{fig:xrayFinger}
\end{figure}

The exact knowledge of the position of the spot on the detector is crucial for the data analysis of the experiment, as it defines the area where the expected signal from the axions is to be focused. For this reason, an Amptek Cool-X x-ray generator was installed on the side of the bore opposite of the XRT-Micromegas line (i.e. sunset side of the magnet), with a mechanism that can move the source in, placing it on the optical axis of the system, and out of the field of view of the optic (referred to as ``x-ray finger''). Once an operational vacuum level was achieved, the Cool-X was turned on to illuminate the x-ray optic and a long integration was acquired. Figure~\ref{fig:xrayFinger} compares the detector data (left panel) with the ray-tracing simulations (center panel) and then overlays an intensity contour map generated from the ray-tracing results on top the data (right panel). There is excellent agreement. 

\begin{figure}[htb]
\begin{center}
\centering
\includegraphics[height=0.32\textwidth]{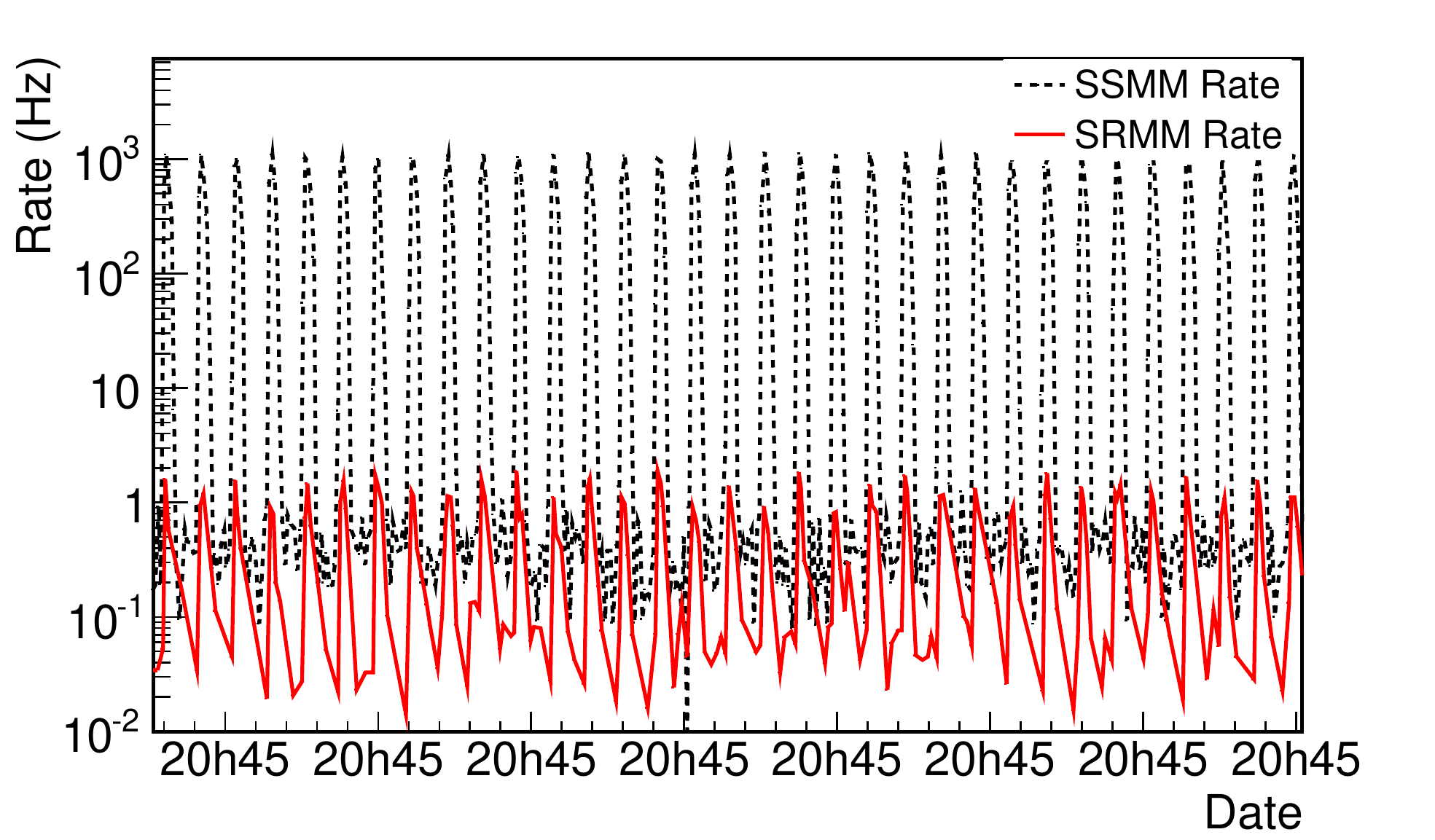} 
\end{center}
\caption{Right: Trigger rate in the sunrise Micromegas and the sunset Micromegas, during an x-ray finger run.}
\label{fig:cal}
\end{figure}

The spot position can be periodically monitored performing spot-calibrations with the Cool-X which emits mainly 8~keV photons and bremsstrahlung x-rays.  The generator does not produce a constant flux of x-rays but is thermally cycled between 2 to 5 minutes; the flux can vary throughout one cycle and from cycle to cycle. Figure \ref{fig:cal} shows the rate recorded by the Micromegas located at the opposite side of the bore, looking for axions during the sunset, compared to the rate of the sunrise detector, sitting at the back of the x-ray generator. The cycles of the generator are evident as well as the 3 orders-magnitude of difference in the recorded rates.

\begin{figure}
\centering
\includegraphics[height=0.35\textwidth]{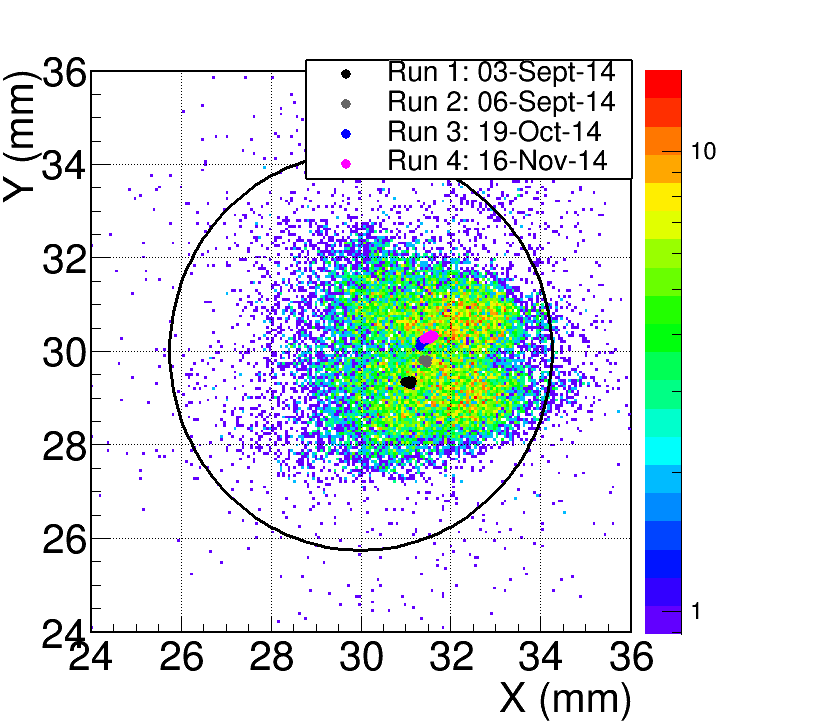} 
\includegraphics[height=0.35\textwidth]{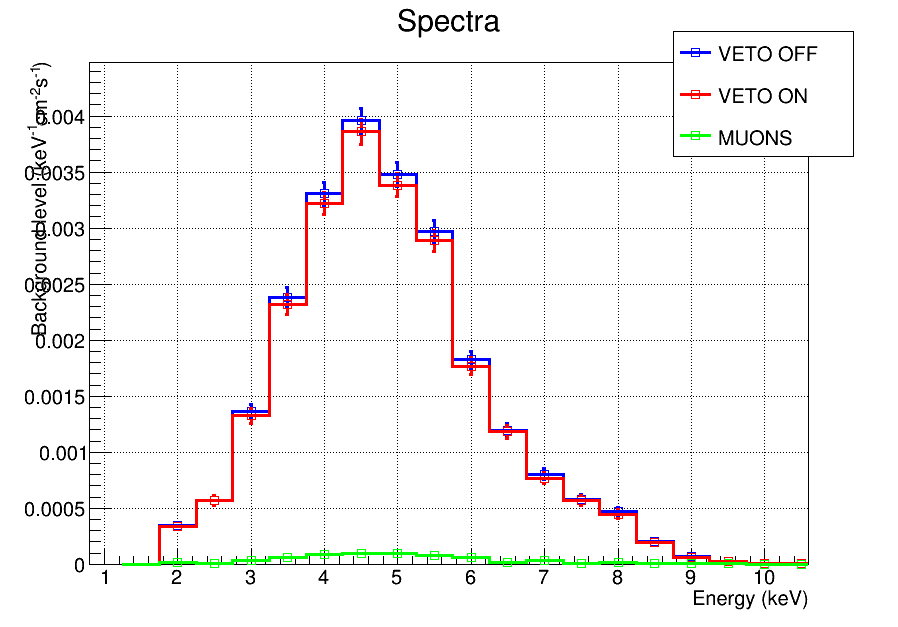} 
\caption{Left: Intensity map of the calibration with the x-ray finger through the x-ray optic. The outer contour is the region used (conservatively) to define the background in the spot. Each of the four dots corresponds to the centroid of a spot-calibration with the x-ray finger. Their statistical error are within the dot size. Right: Energy spectrum of an x-ray finger run. The higher end of the spectrum is reduced because of the focusing optic's efficiency loss at these energies. }
\label{fig:finger}
\end{figure}

As the source is located at a finite distance to the optic, the spot on the detector plane has a larger diameter than the expected signal (see Section \ref{sec:performance}). A spot-calibration can be seen in Figure \ref{fig:finger}. An intensity map is shown on the left, while on the right, the energy spectrum acquired during such a calibration is plotted. The higher end of the spectrum is suppressed due to the drop in efficiency of the optic at these energies.

Reference runs with the x-ray finger were performed just after the alignment of the system, before and after the sunrise detector shielding was in place. The low count rate from the source at such a distance enforces runs of approximately 8$\;$h in order to collect data with significant statistics. For this reason, spot-calibration runs cannot be taken very often so as not to disturb the normal data-taking program of the experiment. Two more spot-calibration runs were performed, one after the installation of the sunset Micromegas detectors and their shielding at the opposite end of the magnet and one after the data taking period, in mid-November 2014. The centroid of the observed spot events has been used to evaluate any possible changes of the spot position during the period of the 2014 data-taking. Table \ref{tab:spot} summarizes the measurements. A small shift is observed when adding the shielding around the detector (second calibration) and the shielding around the sunset detectors at the opposite side of the magnet (third calibration). However, only a a shift of less than 0.2$\;$mm is observed at either direction (x or y) between the last two spot-calibrations, which represent the positions just before and just after the 2014 data-taking of the experiment.

\begin{table}[!ht]
\begin{center}
\begin{tabular}{ccccccc}
\multirow{2}{*}{Year}	&  \multirow{2}{*}{Run} &  Time & \multirow{2}{*}{Events} & \multirow{2}{*}{Conditions} & $x$-centroid & $y$-centroid \\
	&   &  (hours) &  &  & (mm) & (mm) \\
\hline
\hline\\[-3mm]
\multirow{4}{*}{2014}	& 0 & 9.13   & 8910 & No shielding around detectors 	& 31.09 $\pm$ 0.01 &  29.35 $\pm$ 0.01 \\
			& 1 & 5.29   & 4287 & Shielding around sunrise MM  	& 31.46 $\pm$ 0.01  & 29.80 $\pm$ 0.01 \\
			& 2 & 12.43  & 8617 & Shielding around sunset MM 	& 31.34 $\pm$ 0.01 & 30.21 $\pm$ 0.01 \\
			& 3 & 13.94  & 7505 & End of 2014 data-taking 	& 31.53 $\pm$ 0.01 &  30.30 $\pm$ 0.01 \\[1mm]
\hline\\
\end{tabular}
\end{center}
\vspace{-4mm}
\caption{Details of the four x-ray finger runs taken in 2014 and the corresponding spot positions. The third run corresponds to the beginning of the data-taking period for 2014. Some change in position is observed with each mounting of the shielding at the two ends of the magnet, however the position of the spot before and after the data-taking did not differ significantly.}
\label{tab:spot}
\end{table}

\subsection{Detector performance}

\begin{figure}[htb]
\centering
\includegraphics[height=0.35\textheight]{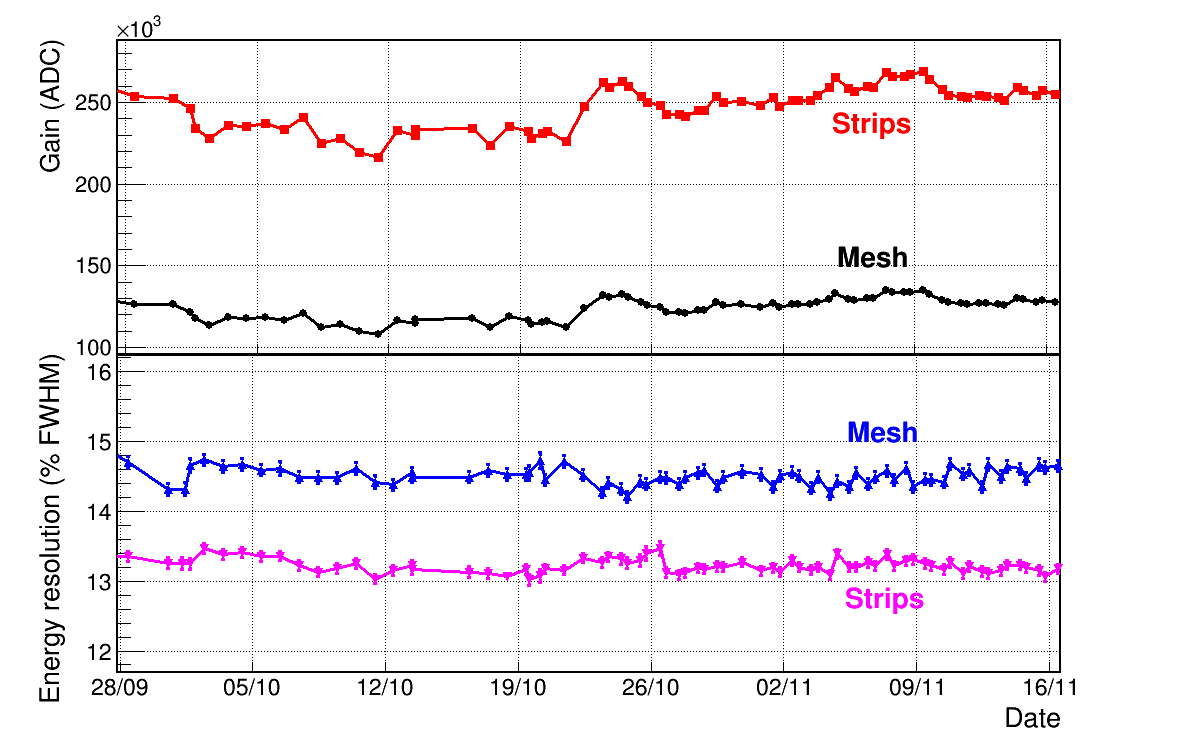} 
\caption{Evolution of the gain and the energy resolution (\%FWHM at 5.9~keV) of the detector during the data-taking. The fluctuations, measured both at the mesh and at the strips are less than 5\%.}
\label{fig:detpar}
\end{figure}

The system performance during the $\sim$50-day data-taking period at the end of 2014 has been stable, as can be seen from the evolution of the monitored parameters like the gain and the energy resolution, plotted in Figure \ref{fig:detpar}. 
Detector calibrations are performed at least twice a day, once after tracking the Sun in the morning and once during background acquisition. For this purpose, an actuator places an $^{55}$Fe source in front of the Micromegas detector and retracts it into a shielded position after the calibration. An imprint of the cathode window strongback can be clearly distinguished during the calibration (Figure \ref{fig:SRTelesStrongBack} right). As already mentioned, the size of the window is calculated to cover an area larger than the expected signal spot, the position of which is indicated by  a solid black line at the center of the plot. 

\begin{figure}[htb]
\centering
\includegraphics[height=0.35\textwidth]{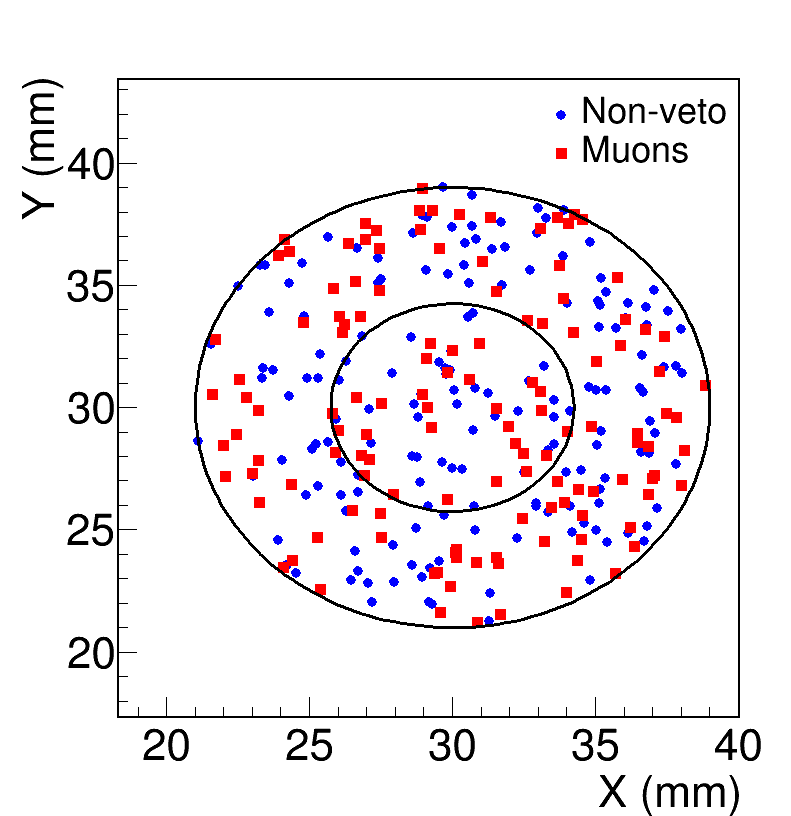} 
\caption{A 2D hitmap of background events in the calibration area (outer circle) with energies between 2 and 14~keV, where the two populations of muon-induced events (muons) and the rest (non-veto) are indicated in red and blue respectively. The inner circle marks the position of the cathode window opening, which includes the spot.}
\label{fig:hitmap}
\end{figure}

The data analysis relies on the study of the calibration events, corresponding to x-ray photons of the energy RoI, and the patterns they induce to several observation parameters. These patterns are then compared to the same parameters of the background events. A detailed description can be found in \cite{Aune:2013nza}. Events are registered uniformly over the full detector surface (see Figure \ref{fig:hitmap}); a good part of them has a temporal stamp that coincides with the signal of the muon-veto, labelled ``Muons'', which are then rejected by the analysis. The background level achieved at a surface covering the calibration area is (1.6~$\pm$~0.2)$\times$10$^{-6}\;$keV$^{-1}$cm$^{-2}$s$^{-1}$, which with the veto condition applied is reduced to (1.0~$\pm$~0.2)$\times$10$^{-6}\;$keV$^{-1}$cm$^{-2}$s$^{-1}$. 


\section{Conclusions and prospects} \label{sec:conclusion}

We have successfully built and commissioned a new low-background x-ray detection line, composed of a cone-approximation Wolter I x-ray telescope coupled to a shielded Micromegas detector. The telescope is sized to cover the CAST magnet bore ($\sim$5 cm diameter) and has been designed to optimize the signal-to-noise ratio of a detection system in CAST. It is comprised of 13 thermally-formed glass substrates deposited with multilayer coatings. It is the first time an x-ray optic is conceived and built specifically for axion physics. The Micromegas detector is placed at the focal point of the telescope, at a distance of 1.5~m, and is able to image the few mm$^2$ focal spot expected from solar axions.

The system has been installed and commissioned in one of the two sunrise exits of the CAST magnet in August 2014 and is currently taking data. The installation, commissioning and first calibration data have been described in detail. The expected specifications, in terms of spot size and throughput, agree with first experimental data. The background level of the detector has once more improved over previous designs, reaching a value of (1.0~$\pm$~0.2)$\times$10$^{-6}\;$keV$^{-1}$cm$^{-2}$s$^{-1}$, the lowest achieved at CAST so far; this value corresponds to 5.4$\times$10$^{-3}\;$counts per hour in the energy RoI and signal spot area. This performance translates to the best signal-to-noise ratio by far obtained by any detection system of the CAST experiment, and will contribute to push the sensitivity of the experiment for the ongoing data taking campaign in 2015.

The system has been conceived as a technological pathfinder for the future International Axion Observatory (IAXO), as it combines two of the techniques (optic and detector) proposed in the conceptual design of the project. It is the first time these two elements are combined and operated together in real data-taking conditions. The operational experience here described  is therefore an important milestone in the preparatory phase of IAXO.

\acknowledgments
We thank our colleagues at CAST for many years of collaborative work and R. de Oliveira and his team at CERN for the manufacturing of the microbulk readouts. We also thank D. Calvet from CEA/Saclay for his help with the AFTER electronics. We acknowledge the support from the European Commission under the European Research Council T-REX Starting Grant ref. ERC-2009-StG-240054 of the IDEAS program of the 7th EU Framework Program. We also acknowledge support from the Spanish Ministry MINECO under contracts ref. FPA2008-03456 and FPA2011-24058, as well as under the CPAN project ref. CSD2007-00042 from the Consolider-Ingenio 2010 program. These grants are partially funded by the European Regional Development funded (ERDF/FEDER). Part of this work was performed under the auspices of the U.S. Department of Energy by Lawrence Livermore National Laboratory under Contract DE-AC52-07NA27344. 
F.I. acknowledges the support from the \emph{Juan de la Cierva} program and T.D. from the \emph{Ram\'on y Cajal} program of MICINN.

\bibliographystyle{jhep}
\bibliography{biblio3}

\end{document}